\documentstyle[epsf]{mn}
\newcommand{\be}{\begin{equation}}
\newcommand{\ee}{\end{equation}}
\title[Inverting the lens] 
{Non-parametric inversion of strong lensing systems.}  
\author[Diego et al.]  
  {J.M. Diego$^1$, P. Protopapas$^2$, H.B. Sandvik$^3$, M. Tegmark$^4$.\\  
   University of Pennsylvania. 209S, 33rd St,   
   Department of Physics \& Astronomy, Philadelphia PA 19104,
   USA.\\$^1$jdiego@physics.upenn.edu \\$^2$pavlos@physics.upenn.edu \\$^3$sandvik@hep.upenn.edu \\$^4$max@hep.upenn.edu}   
\date{Draft version \today}  
  
\pagerange{\pageref{firstpage}--\pageref{lastpage}}  
  
\begin{document}  
\maketitle  
 
\label{firstpage}  
%%%%%%%%%%%%%%%%%%%%%%%%%%%%%%%%%%%%%%%%%%%%%%%%%%%%%%%%%%%%%%%%%%%%%%%%%%%%%%%  
\begin{abstract}  
We revisit the issue of non-parametric gravitational lens reconstruction  
and present a new method to obtain the cluster mass distribution using   
strong lensing data without using any prior information on the  
underlying mass.   
The method relies on the decomposition of the lens plane into individual  
cells. We show how the problem in this approximation can be   
expressed as a  system of linear equations for which a   
solution can be found.  Moreover, we propose to include information  
about the {\it null space}.  That is, make use of the pixels where we  
know there are no arcs above the sky noise.   
The only prior information is an estimation of the physical size of   
the sources. No priors on the luminosity of the cluster or shape of the halos   
are needed thus making the results very robust. In order to  
test the accuracy and bias of the method we make use of simulated strong   
lensing data.  
We find that the method reproduces accurately both the lens mass and  
source positions and provide error estimates. 
\end{abstract}  
%%%%%%%%%%%%%%%%%%%%%%%%%%%%%%%%%%%%%%%%%%%%%%%%%%%%%%%%%%%%%%%%%%%%%%%%%%%%%%%  
%%%%%%%%%%%%%%%%%%%%%%%%%%%%%%%%%%%%%%%%%%%%%%%%%%%%%%%%%%%%%%%%%%%%%%%%%%%%%%%  

%%%%%%%%%%%%%%%%%%%%%%%%%%%%%%%%%%%%%%%%%%%%%%%%%%%  
\section{Introduction}\label{section_introduction}  
%%%%%%%%%%%%%%%%%%%%%%%%%%%%%%%%%%%%%%%%%%%%%%%%%%%  
Analysis of strong lensing images in galaxy clusters is one of the 
more subjective and intuitive fields of modern astronomy 
(see Blandford \& Narayan 1992, Schneider, Ehlers \& Falco  1993, 
Wambsganss 1998, Narayan \& Bartelmann 1999, Kneib 2002, for comprehensive 
reviews on gravitational lensing).  
The common use of parametric models means making educated choices  
about the cluster mass distribution, for instance that the dark matter 
haloes follow the luminous matter in the cluster or that galaxy profiles 
posses certain symmetries. Once these choices are 
made, one can only hope they are the right ones.  
The recent improvement in strong lensing data however, is impressive
and should increasingly
allow us to extract information using fewer assumptions. For
instance, it should allow us to relax assumptions about the mass
distributions of the cluster and instead enable us to test this. The
approach of testing rather than assuming the underlying physics has
now been used in other branches of astrophysics and cosmology
(e.g. Tegmark 2002) but has so far been largely absent in strong
lensing analyses.
  
There are two relatively different fields within the area of strong  
lensing, namely lensing by galaxies and galaxy clusters. These are  
different both in appearance as well as in abundance. Galaxy-cluster  
systems are few and far between, only the most abnormally dense  
clusters have surface densities greater than the critical density for  
lensing. However, they are much more spectacular than  
their galaxy lens counterparts. The most massive clusters are able to  
create multiple images of extended objects such as distant galaxies  
with image separations of up to $\sim 1$ arcminute. Arguably the most  
impressive system, A1689 (Broadhurst et al. 2005), boasts more than 100 multiple images of  
some 30+ sources. Galaxy lens systems are of course on a much smaller  
scale with image separations of a few arcseconds.   
They are far more abundant, but less impressive.   
The lensed objects that we are able to  
observe are typically high redshift quasars due to their high  
luminosity and point like structure, although galaxy-galaxy lensing has  
been observed ( Brainerd, Blandford \& Smail 1996, 
Hudson et al. 1998, Guzik \& Seljak 2002).   

The classification of  
strong lensing systems into these two groups is not merely a matter of  
scale. The baryons in galaxies have had time to cool and form the  
visible galaxy, thereby giving a cuspy profile, suitable for  
lensing. The cooling time for galaxy clusters exceed the Hubble time  
and the density profile is therefore far less cuspy, making them less  
ideal lenses. Although cluster lens systems are scarce, the impressive  
number of lensed images in each system, mean they still contain a lot of  
information, particularly regarding cluster mass profiles. The  
information is harder to extract due to the  
relatively complicated gravitational potentials, but if this challenge  
can be overcome they could be highly useful probes, certainly of cluster  
physics and potentially also for cosmology 
(Yamamoto \& Futamase 2001, Yamamoto et al. 2001, Chiba \& Takahashi 2002, 
Golse et al. 2002,  Sereno 2002, Meneghetti et al 2005)
The intention of this paper is to improve our methods for extracting this 
information.

Although alternative approaches have been suggested to recover   
the density field in both the weak and strong lensing regime, 
(see for instance Kaiser \& Squires 1993,   
Broadhurst et al. 1995, Kaiser 1995, Schneider 1995,    
Schneider \& Seitz 1995, Seitz \& Schneider 1995,   
Bartelmann et al. 1996, Taylor et al. 1998, Tyson et al. 1998, 
Bridle et al. 1998, Marshall et al. 1998),   
the standard approach to modeling strong cluster lenses is using  
parametric methods. This is motivated   
by the fact that the data usually do not contain more than a few arcs.   
This is not enough to constrain the mass   
distribution without the help of a parametrization. Parametric methods   
rely heavily on assumptions or priors on the mass distribution 
(Kochanek \& Blandford 1991, Kneib et al. 1993, 1995, 1996, 2003,   
Colley et al. 1996, Tyson et al. 1998, Broadhurst et al. 2000, 2005,  
Sand et al. 2002). A common prior is the assumption that there is a 
smooth dark matter component which is correlated spatially with the 
centroid of the luminous matter in the cluster. The mass is then ususally 
modeled by a large smooth dark matter halo placed on top of the central galaxy or the 
centroid of the luminous matter, as well as smaller dark matter haloes located in 
the positions of the other luminous galaxies. The parameters of each halo are then 
adjusted to best reproduce the observations.

There is plenty of subjectivity involved in this process, particularly
in the addition of the dominant dark matter  
component to the cluster. The assumption that the dark matter follows 
the luminosity is necessary but remains the Achilles heel of parametric 
lens modelling. For large clusters the number of parameters in the 
parametric lens model quickly becomes large but 
there is still no guarantee that the parametric model used, is in fact 
capable of reproducing well the mass distribution. It is not hard to envisage
complications like dark matter substructure, asymmetric galaxy
profiles, interactions between individual galaxies and the cluster or
even dark matter haloes without significant luminosity all of which
would not be well represented by the typical parametric methods.   
In these cases, where the number of parameters is large, we may want 
to consider alternative non-parametric methods where all the previous 
problems do not have any effect on any of the assumptions. Also 
is in these situations where the number of parameters in both parametric 
and non-parametric methods is comparable. When the number of parameters 
is comparable in both cases, it is interesing to explore non-parametric 
methods since they do not rely on the same assumptions. 
 
This paper does not pretend to be an attack to parametric methods but a 
defense of non-parametric alternatives. Parametric methods are usually  
the best way to obtain information about the gravitational potential 
(few multiple images, simple lens) and are not affeted by resolution problems, 
specially in the center of the lens which can play a key role in reproducing the 
radial arcs. They however suffer of some potential problems, some of which have been 
outlined above. These problems combined with the impressive quality of recent 
and upcoming data lead us in this paper to explore the potential to constrain the mass
distribution without imposing priors on it. In other words we want to
know what strong gravitational lensing images can tell us about
cluster mass profiles whilst pretending we know nothing  about the
luminosity.  
Upcoming images of strong lensing in galaxy clusters will contain of
the order of a hundred arcs and should make this a manageable
task (Diego et al. 2004). A non-parametric approach will provide an important
consistency check, since concurring results would lend strength to the
parametric approach, whereas any resulting differences would need to be
addressed. 
  
Accepting this challenge we present in this paper a new method which 
makes use of all the available information in super-high quality strong  
lensing systems. The method has also been thoroughly tested with
simulated lensing data with very good results.  We thus address the crucial
issue of how well we can reproduce both the cluster mass profile and
the positions and shapes of the background  galaxies.  
  
%%%%%%%%%%%%%%%%%%%%%%%%%%%%%%%%  
%\section{Relation to previous work}  
%%%%%%%%%%%%%%%%%%%%%%%%%%%%%%%%%  
We want to stress that ours is by no means the first work
proposing use of non-parametric methods in strong lensing. Among the
non-parametric methods already available, are
the pixellization methods of  Saha et al. (1997, 2000), Abdelsalam et
al. (1998a, 1998b), Williams et al. (2001), who first established many
of the ideas revisited in this work, as and also the multi-pole
approach of Kochanek \& Blandford (1991), Trotter et al. (2000). 

Naturally, our work shares many similarities with this former work,
but there are also important and interesting differences.  
Firstly, in Saha et al. (1997) (and subsequent papers), the authors divide the   
lens plane into a grid similar to the method presented here, but with
the important difference that their grid is fixed while our grid is
dynamical. Our grid adapts to the new estimated mass at each step. This has
important implications for the solution since high density regions
will be sampled more heavily. These dense regions play a key role,
particularly in the positions of the radial arcs. Not sampling these
regions properly will necessarily lead to a biased mass distribution.

Another important difference with Saha et al (1997) is that they make 
use of a prior on the mass distribution, penalizing deviations
from the distribution of luminous matter.
They claim this prior does not play an important role but we find this
questionable. It is hard to quantify the effect since the method
apparently was not 
tested on simulated lensing data. In contrast, as mentioned above, our
method is thoroughly tested with simulations to quantify how well the mass
distribution is recovered. We do not use any prior other than a
{\it physical} prior on the sizes of the sources. This prior is proved
to be weak provided it is chosen  with a minimum of wisdom.  
 
A third important difference between the work presented in this   
paper and others is that we show how to speed up the algorithm   
significantly by adopting techniques commonly used in optimization 
problems.   
 
Moreover as an added novelty, our algorithm also includes for the
first time information  
about the null space. Rather than using only the information in the  
lensed arcs, we use information which has hitherto been overlooked;  
the areas in the sky where no arcs are observed.

The paper is laid out as follows. In the next section we present the
lens inversion problem, and provide a linear approximation. We then in
section \ref{simulations} present the simulated lensing data used to
test the method, before we go on to present the adaptive gridification
of the lens-plane, section \ref{gridifying}. 
In section \ref{inversion} we describe the various inversion
algorithms and finally, in section \ref{nullspace} we introduce the use
of the complementary (null) space. 

%%%%%%%%%%%%%%%%%%%%%%%%%%%%%%%%%%%%%%%%%%%%%%%%%%%%%%%%%%%%%%%%  
\section{The problem formulated in its basic {\it linear} form}  
\label{linear}  
%%%%%%%%%%%%%%%%%%%%%%%%%%%%%%%%%%%%%%%%%%%%%%%%%%%%%%%%%%%%%%%%  

The fundamental problem in lens modelling is the following: Given the   
positions of lensed images, $\vec{\theta}$, what are the positions of the  
corresponding background galaxies $\vec{\beta}$ and the mass distribution of  
the lens, $M(\vec{\theta})$. Mathematically this entails inverting the lens equation  
\begin{equation}  
\vec{\beta} = \vec{\theta} - \vec{\alpha}(\vec{\theta},M(\vec{\theta}))  
\label{eq_lens}  
\end{equation}  
where $\vec{\alpha}(\vec{\theta})$ is the deflection angle created by the lens   
which depends on the observed positions, $\vec{\theta}$. From now on  
we will omit the vector notation unless otherwise noted. The {\it data points} 
of our problem, $\vec{\theta}$, are given by the $x$ and $y$ positions of each 
of the pixels forming the observed arcs.   
  
The deflection angle, $\alpha$, at the position $\theta$, is found by  
integrating the contributions from the whole mass distribution.  
\begin{equation}  
\alpha(\theta) = \frac{4G}{c^2}\frac{D_{ls}}{D_s D_l} \int M(\theta')  
                 \frac{\theta - \theta'}{|\theta - \theta'|^2} d\theta'  
\label{eq_alpha}  
\end{equation}  
where $D_{ls}$, $D_l$, and $D_s$ are the angular distances  
from the lens to the source galaxy, the distance from the observer to  
the lens and the distance from the observer to the source galaxy  
respectively. In equation    
\ref{eq_alpha} we have made the usual thin lens approximation   
so the mass $M(\theta')$ is the projected mass along the line of   
sight  $\theta'$. Due to the (non-linear) dependency of the deflection angle,  
$\alpha$ on the position in the sky, $\theta$, this  
problem is usually regarded as a typical example of a non-linear  
problem. We will see that this is only partially true.  
  
The next approximation we make is to split the lens   
plane in $N_c$ small regions (hereafter cells) over which the projected mass   
is more or less constant. We can then rewrite equation (\ref{eq_alpha}) as;  
\begin{equation}  
\alpha(\theta) = \frac{4G}{c^2}\frac{D_{ls}}{D_s D_l} \sum_{N_c} m_i  
                 \frac{\theta - \theta_i}{|\theta - \theta_i|^2}   
\label{eq_alpha2}  
\end{equation}  
The first point we want to make here is that the deflection angle   
$\alpha$ may be thought of as the net contribution of many small masses $m_i$ in   
the positions $\theta_i$, each one pulling the deflection in the   
direction of ($\theta - \theta_i$) and with a magnitude which is proportional   
to $m_i/(\theta - \theta_i)$. If we divide the lens plane in a grid   
with $N_c$ cells, the masses $m_i$ can be considered as the mass   
contained in the cell $i$ ($i = 1, ..., N_c$). If the cells  
are sufficiently small then the above   
pixellization of the mass plane will give a good approximation to the real   
mass distribution.  
  
Our second point is that the problem is non-linear in   
one direction but linear in the other. That is, given a position   
in the sky $\theta$ (and given a lens) there is only one $\beta$ which   
satisfy equation \ref{eq_lens} but given a position of the background   
galaxy $\beta$, there may be more than one position in the sky ($\theta$)  
satisfying equation \ref{eq_lens} or equivalently, the source galaxy   
may appear lensed in more than one position in the sky. The linear nature of   
the problem is evident when one realizes that the only non-linear variable,   
namely the $\theta$ positions, are fixed by the observation and that the problem    
depends linearly on the unknowns ($\beta$ positions and masses, $m_i$, in the cells).  
  
Let us now assume that we have a data set consisting of a system   
of radial and tangential strong lensing arcs which are spread over $N_{\theta}$   
pixels in the image. We will also assume that we know which arcs  
originate from the same sources. Since both the data and the mass distribution has  
been discretized we can rewrite equation (\ref{eq_lens}) as a  
simple matrix equation:  
\begin{equation}  
\beta = \theta - \Upsilon M  
\label{eq_lens_matrix}  
\end{equation}  
where $\theta$ and $\beta$ are now $2N_{\theta}$ element vectors  
containing $x$ and $y$ values of the observed positions and the (unknown)  
source positions respectively.  
$M$ is the mass vector containing all $N_c$ mass cells, and $\Upsilon$  
is the ($2N_{\theta} \times N_c$)  matrix casting the mass vector into  
a vector of displacement angles. A more technical account of the  
make-up of the $\Upsilon$ matrix can be found in the appendix.  
\begin{figure}  
   \epsfysize=8.cm   
%   \begin{minipage}{\epsfysize}\epsffile{MassNFWInv.ps}\end{minipage}  
   \begin{minipage}{\epsfysize}\epsffile{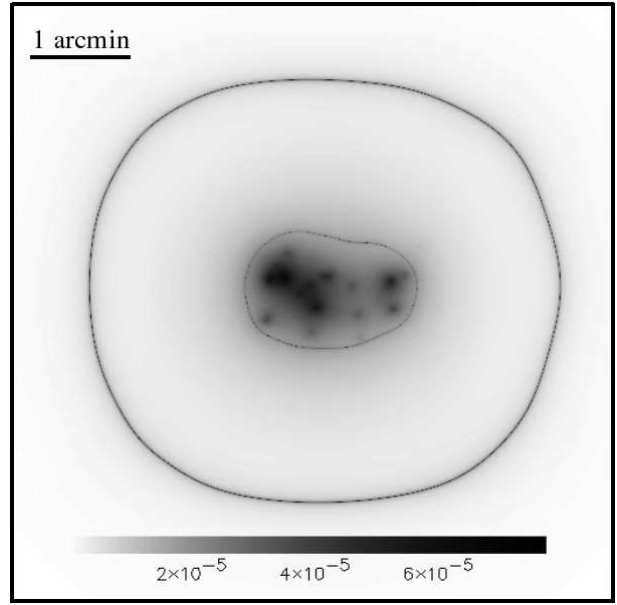}\end{minipage}  
   \caption{  
            Original simulated mass profile. The total projected mass is   
            $1.119 \times 10^{15}$ h$^{-1}$ M$_{\odot}$ in the   
            field of view (0.1 degrees across)    
            and the cluster is at $z=0.18$. The units of the grey scale map 
            are $10^{15} h^{-1} M_{\odot}/$pixel where each pixel is 0.494
            arcsec$^2$. Also shown are the radial (small) and tangential 
            (large) critical curves for a source at redshift $z=3$.
           }  
   \label{fig_mass}  
\end{figure}  
\begin{figure}  
   \epsfysize=8.cm   
%   \begin{minipage}{\epsfysize}\epsffile{SourcesHUDF.ps}\end{minipage}  
   \begin{minipage}{\epsfysize}\epsffile{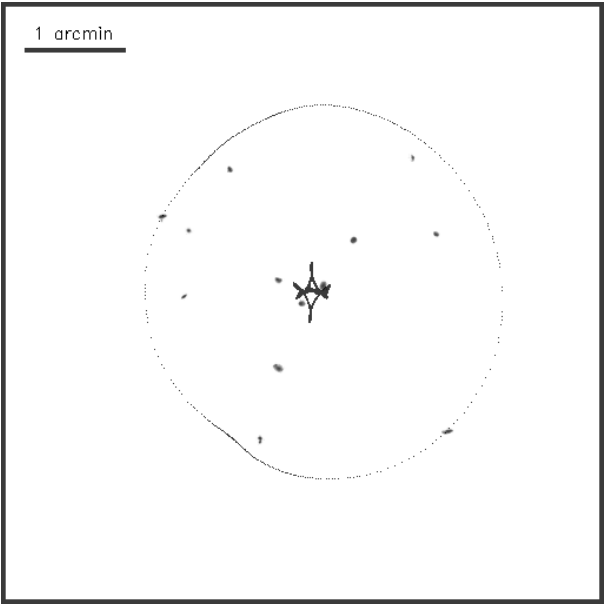}\end{minipage}  
   \caption{  
            The 13 sources at redshift $z \in (1,6)$. The field of view is 
            0.1 degrees across and is centered in the same point 
            as figure \ref{fig_mass}. The solid line marcs the position of the caustics 
            at $z=3$ correponding to the radial (dotted line) and tangential (solid line) 
            critical curves. 
           }  
   \label{fig_sources}  
\end{figure}  
Equation \ref{eq_lens_matrix} clearly demonstrates the linear nature of the problem  
when formulated in this manner. The problem has now been reduced to a set of  
$2N_{\theta}$ linear equations with $2N_{\theta}+N_c$ unknowns (lens masses  
and source galaxy positions).   
Notice that when the problem is formulated in this form, there are more   
unknowns than equations which means it is an underdetermined system  
with an infinite number of solutions. In order to identify a suitable  
solution for such a system, we need to add extra information or impose  
constraints.  
  
One way of doing this is by reducing the number of unknowns, for  
instance by removing the source positions from the unknown category.  
This can be achieved by minimizing the dispersion in the source plane,  
i.e. demanding that the pixels in the source plane be as concentrated as possible for each  
source. In this case we are left with only $N_c$ unknowns.   
  
Another way of constraining the system is by assuming the $N_s$  
sources are well approximated by point sources, which reduces the  
number of unknowns to $N_c + 2N_s$.   
This means effectively demanding that all observed $\theta$s for arcs  
corresponding to the same source, can be traced back through the lens  
to $N_s$ single points. With this assumption we can rewrite the  
lens equation in the compact form of   
\begin{equation}  
\theta = \Gamma X.  
\label{eq_lens2}  
\end{equation}  
$\Gamma$ is now a matrix of dimension $2N_{\theta}\times (N_c+2N_s)$   
and $X$ is the vector of dimension $(N_c+2N_s)$ containing all the   
unknowns in our problem (see appendix A), the mass elements and the $2N_s$   
central ($x$- and $y$) coordinates of the $N_s$ sources. Now the  
system is matematically overdetermined (more data points that unknowns) 
and has a unique \emph{point source solution}.
This unique solution can be found by numerical methods as we will see later.

The linearization of the problem means that it is in principle  
solvable by both matrix inversion and simple linear programming  
routines. In practice, the problem quickly becomes ill-conditioned  
and too large for direct matrix inversion, and approximate numerical  
methods are more suitable. The main problem with the linearization is  
that we do not know if the obtained linearized solution creates  
artificial tangential or radial arcs. Checking this requires forward  
solving of the lens equation which is non-linear due to the  
complicated dependence of the deflection angle on $\theta$.  
  
We suggest a novel approach to this problem, by using  
all the available information in the images, i.e. the  
information inherent in the {\it dark areas} (pixels containing no 
arcs) as well as the observed arcs. 
By pixellization the dark areas, tracing these pixels back through  
the lens and imposing that they fall outside the sources  
it is possible to find the true solution without over-predicting  
arcs. This use of the null space is to our knowledge  
unprecedented, and in principle allows for a complete, linear solution  
to a problem usually considered non-linear.  
  
%%%%%%%%%%%%%%%%%%%%%  
\section{Simulations}  \label{simulations}
%%%%%%%%%%%%%%%%%%%%%  
Before proceeding to invert the system of linear equations (\ref{eq_lens2}),   
it is instructive to take a closer look at the simulations which are  
going to be used to test the inversion algorithms.  
Given that most methods rely on the luminous mass distribution, this is particularly  
important should the galaxies not accurately trace the dark matter.  
  
Our simulations consist of three elements. The first element of the  
simulation is the projected mass distribution ($M$) in the lens 
plane. We simulate a generic mass distribution    
of a cluster with a total projected mass of  $1.119 \times 10^{15}$   
h$^{-1}$ M$_{\odot}$ in the field of view located at redshift $z=0.18$. The field of view of   
our simulation is 0.1 degrees 
The mass profile is built from a superposition of 20 NFW profiles   
with added ellipticity. The halos' masses vary   
from $0.25\times 10^{15}$ h$^{-1}$ M$_{\odot}$   
to $2\times 10^{12}$ h$^{-1}$ M$_{\odot}$.  
In the context of lensing, NFW halos seem to reproduce well the shear   
profile of massive clusters up to several Mpc (Dahle et al. 2003, 
Kneib et al. 2003, Broadhurst et al. 2005).  
The final mass distribution is shown in figure \ref{fig_mass}. 
Also in the same figure we show the two critical curves. 
The interior one is the 
radial critical curve and the exterior one the tangential critical curve.
Both curves have been calculated assuming the source is at redshift $z=3$. 
The tangential critical curve is usually associated with the Einstein radius. 
Its large size (radius $\approx 2$ arcmin) is due to the unusually high 
projected mass of our cluster plus its relatively low redshift ($z=0.18$).
Although the deduced Einstein radius is larger than the one observed in
clusters, the simulation presented here will serve the purpose of testing
the different methods discussed here. More realistic simulations with
proper mass distribution and source density will be presented in a future
paper. 

The second element in our simulation are the sources ($\beta$),   
for which we extract 13 sources from the HUDF   
(Hubble Ultra Deep Field, Beckwith et al. 2003). 
We assign them a redshift ($z \in [1,6]$) and    
size  and place them in 
different positions behind the cluster plane.  
 
\begin{figure*}  
   \epsfysize=8.5cm   
%   \begin{minipage}{\epsfysize}\epsffile{LensedHUDF.ps}\end{minipage}  
%%   \begin{center}{\epsfysize}\epsffile{LensedHUDF.ps}\end{center}  
   \begin{minipage}{\epsfysize}\epsffile{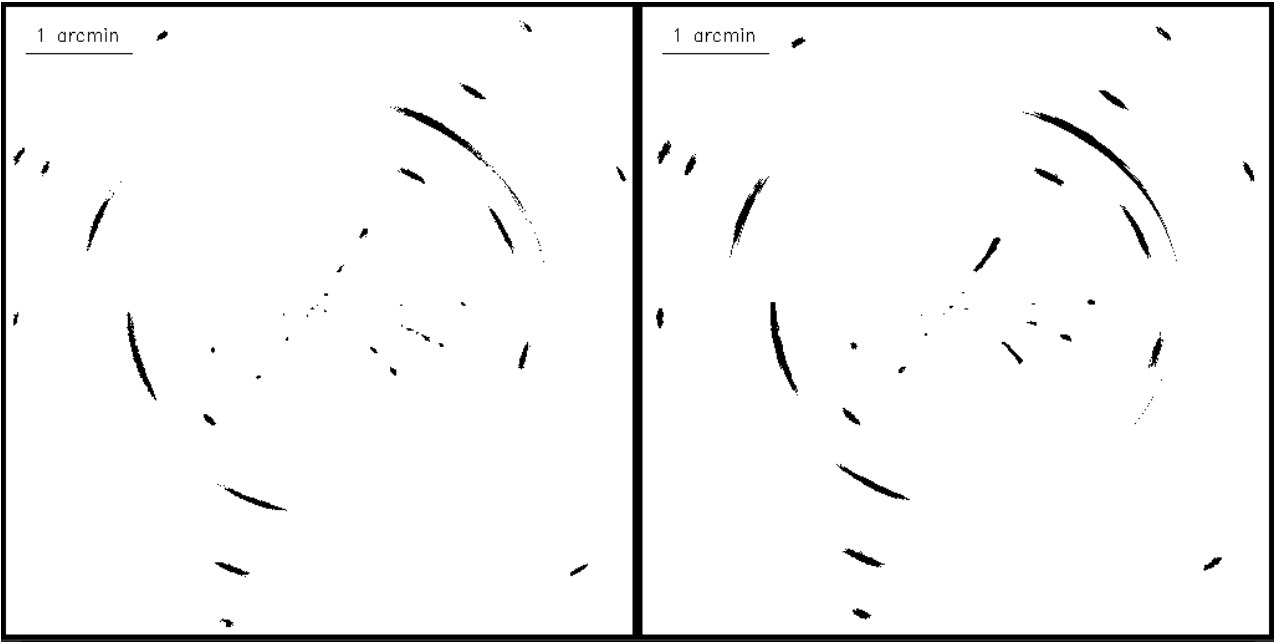}\end{minipage}  
   \caption{  
            The sources lensed by the mass distribution of figure   
            \ref{fig_mass}. The right image is the lensed image using 325   
             cells in the dynamical grid (see text). The left image   
             is the exact solution using no griding. Note the differences   
             in the radial arcs.    
           }  
   \label{fig_theta}  
\end{figure*}

The third element are the lensed images ($\theta$)   
which are calculated from the first two elements ($M$ and $\beta$).   
This is done through a simple ray-tracing procedure. For each position  
$\theta$ in the image, we calculate the deflection angle, $\alpha$,  
and then the corresponding source plane position,$\beta$, according to  
the lens equation (equation \ref{eq_lens}).    
If the calculated $\beta$ coincides with one of the original sources,  
we assign to the lensed image the value (colour) of that  
source. Otherwise that point in the lensed image is left dark (value 0 ).   
We repeat the operation for all the pixels ($\theta$s) to produce a 
complete image. Also, since the sources are small compared with our pixel size and to   
avoid missing some sources, we oversample our $\theta$ pixels by subdividing them   
and checking each pixel at different locations ($\theta + \Delta \theta$ with   
$\Delta \theta < 1$ and $\theta \in [1,512]$)   
The original sources are plotted in figure \ref{fig_sources} and the   
corresponding $\theta$s in the left panel of figure \ref{fig_theta}.  
  
%%%%%%%%%%%%%%%%%%%%%%%%%%%%%%%%%%%%%%%%%%  
\section{Gridifying the mass distribution}  \label{gridifying}
%%%%%%%%%%%%%%%%%%%%%%%%%%%%%%%%%%%%%%%%%%  
\subsection{The $\Upsilon$ matrix}  
As laid out in section \ref{linear} the basis of our non-parametric  
reconstruction method is the assumption that the real mass can be well  
approximated by a pixelized mass distribution. The $\Upsilon$ matrix is  
then the matrix which casts the mass vector into the vector of  
deflection angles  
\be  
\alpha_i = \Upsilon_{ij}M_j  
\ee  
  
For convenience rather than taking the mass distribution in each cell 
to be constant we assume it follows a Gaussian distribution centred in 
the  centre of the cell with some dispersion. This allows us to 
calculate analytically the lensing contribution from each mass cell, 
saving valuable computer time. We use a dispersion of $2a$ where $a$ 
is the size of the cell, and we have confirmed with simulations that 
the lensed image generated with this choice agrees well with the true 
lensed image using a relatively small number of cells. The detailed 
structure of the $\Upsilon$ matrix will be explained in appendix A.   
  
%%%%%%%%%%%%%%%%%%%%%%%%%%%%%%%%%%%%%%%%%%%%%%%%%%%%%%%%%  
\subsection{Multi resolution mass-grid}\label{sect_grid} 
%%%%%%%%%%%%%%%%%%%%%%%%%%%%%%%%%%%%%%%%%%%%%%%%%%%%%%%%% 
Rather than taking a uniform grid, it is better to construct a   
{\it dynamical}- or {\it multi-resolution} grid.  
By sampling dense regions more heavily, it is 
possible to reduce drastically the number of cells needed to
accurately reproduce 
the lensing properties of the cluster. In other 
words we choose an adaptive grid which samples the dense cluster centre 
better than the outer regions. Since we do not actually know the 
density profile of the cluster, this multi-resolution grid must be 
obtained through an iterative procedure. An example of this 
process can be found on the 
SLAP\footnote{see http://darwin.cfa.harvard.edu/SLAP/} 
(Strong Lensing Analysis Package) webpage.  
 
Given a mass estimate (a first mass estimate can be obtained with   
a coarse regular grid), we split a given cell into four sub-cells if the 
mass in the cell exceeds some threshold value. The lower this 
threshold, the higher the number of divisions and consequently the 
higher the final number of cells. The obtained grid can then be used for the 
next mass estimate, and the process can be repeated as 
necessary. Typically the mass estimate will improve with each 
iterative step as this dynamical 
grid allows for the relevant regions of the cluster to become resolved.  
 
\begin{figure}  
   \epsfysize=8.0cm   
%   \begin{minipage}{\epsfysize}\epsffile{DynGrid.ps}\end{minipage}  
   \begin{minipage}{\epsfysize}\epsffile{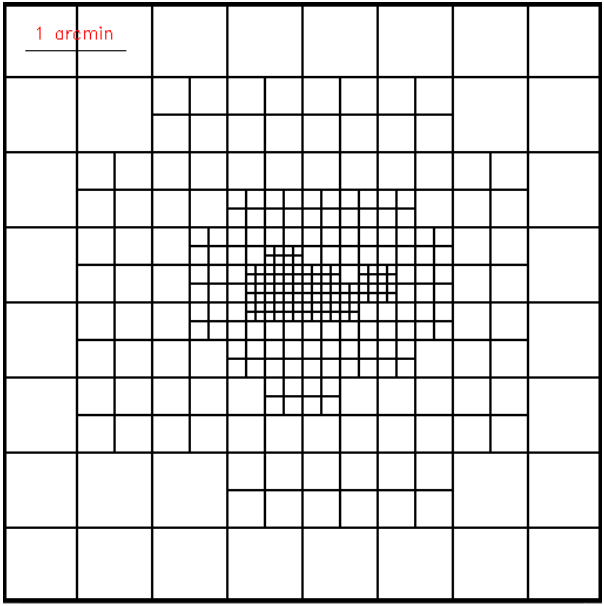}\end{minipage}  
   \caption{  
            Dynamical grid for the mass in figure \ref{fig_mass}.  
            This case corresponds to $M_{thr} = 8.0 \times 10^{12}$   
            h$^{-1}$ M$_{\odot}$.  
           }  
   \label{fig_dynamicgrid}  
\end{figure}

\begin{figure}  
   \epsfysize=8.0cm   
%   \begin{minipage}{\epsfysize}\epsffile{DynGridMassInv.ps}\end{minipage}  
   \begin{minipage}{\epsfysize}\epsffile{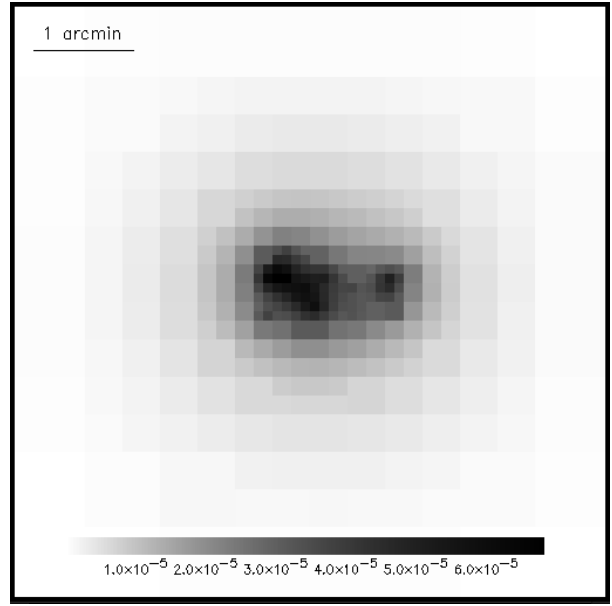}\end{minipage}  
   \caption{  
            Masses in the cells of the dynamical grid of figure \ref{fig_dynamicgrid}.   
           }  
   \label{fig_massgrid}  
\end{figure}

In figure \ref{fig_dynamicgrid} we show an example of a gridded version of the   
true mass in figure \ref{fig_mass} for a threshold of   
$M_{thr} = 8.0 \times 10^{12}$ h$^{-1}$ M$_{\odot}$.   
This grid has 325 cells.   
The corresponding (true) mass in the grid is shown in figure \ref{fig_massgrid}. 
The parameter $M_{thr}$ or equivalently the number of cells, $N_c$, can be viewed 
as a free parameter but also as a prior. Fixing it means we are fixing the minimum scale 
or mass we are sensitive too. In a paralell paper (Diego et al. 2004) we explored 
the role of $N_c$ and found that it can have a negative effect on the results 
if it takes very large values. A natural upper limit for $N_c$ is 2 times the number of 
pixels in the $\theta$ vector minus 2 times the number of sources (the factor 2 accounts 
for the $x$ and $y$ positions). Our simulations show that a good election 
for $N_c$ is normally $1/4$ of this upper limit.\\

For the shake of clarity it is useful to give some practical numbers. The image 
on the left of figure \ref{fig_theta} has $512^2$ pixels from which 2156 pixels 
are part of one of the $\approx 35$ arcs so $N_{\theta} = 2156$. These arcs are coming 
from 13 sources ($N_s = 13$) and the lens plane is tipically divided in a few hundred cells 
($N_c \sim 300$).

%%%%%%%%%%%%%%%%%%%%%%%%%%%   
\section{Inversion methods} \label{inversion}
%%%%%%%%%%%%%%%%%%%%%%%%%%%  
In this section we will describe some inversion methods which can be   
applied to solve the problem. Most of these methods can be found in   
popular books like {\it Numerical Recipes} (Press et al. 1997).   
All these algorithms are being implemented in the package SLAP$^{\star}$   
%http://darwin.cfa.harvard.edu/SLAP})  
which will be made available soon. 
  
Once we have the problem formulated in its linear form with all the unknowns   
on one side it is tempting to try a direct inversion of equation   
\ref{eq_lens2}. Although the $\Gamma$ matrix is not square, one can find   
its inverse, $\Gamma^{-1}$, by decomposing $\Gamma$ into orthogonal matrices.   
This is similar to finding the {\it eigenvalues} and {\it eigenvectors} of   
$\Gamma$. This approximation has its advantages as well as its
drawbacks and we   
will explore this possibility later. However, we anticipate that
degeneracies between neighboring   
pixels in the arcs as well as neighboring cells in the lens plane (not to mention   
the compact nature of the sources)  will result in a system of linear
equations which is not well behaved.
The rank of the matrix $\Gamma$ will   
be normally smaller than its smaller dimension. Calculating the inverse in   
this situation is not a trivial task. 
  
A second approach is {\it rotating} our system of linear equations using   
a transformation which is just $\Gamma^T$. This transforms $\Gamma$ into a square,   
symmetric and positive definite matrix of dimension $(2N_s + N_c)\times(2N_s + N_c)$,   
$A = \Gamma^T\Gamma$ which is better behaved than the original $\Gamma$ matrix.
However, the rank of $A$ is in generally smaller than its dimension and   
its inverse does not exist. The hope in this case must therefore be to 
find an approximate rather than exact solution to the system.  

The third approach is the simplest and will be explored first. We assume   
we know nothing about the sources other than their redshift and that they are much    
smaller than the strong lensing arcs. This is the same as saying that   
the lens has to be such that it {\it focuses} the arcs into compact sources   
at the desired (known) redshift. This simple argument alone will turn out to be   
powerful enough to get a quick but good first estimate of the mass 
distribution in the lens plane.  
   
We explore all these approaches below in reverse order.   
  
%------------------------------------------------------------------------------  
\subsection{A first approach: Minimizing  
  dispersion in the source plane} \label{minvariance}  
%-----------------------------------------------------------------------------  
In this subsection we will discuss the simplest (although  
effective) method to get a fast estimation of the mass using no   
prior information on the lens nor the sources.  
The problem then contains two sets of unknowns: The mass vector   
we want to determine, $M$, and the $\beta$ positions of the   
sources. In general, for a finite number of observed arcs,   
there are several combinations of $\beta$ and $M$ which can   
reproduce the observations. The most obvious unphysical solution is 
the null solution, where the mass is zero and the sources identical to 
the observed arcs.  
The easiest way to avoid such unphysical solutions is to minimize the 
variance of the $\beta$ positions. 
This is equivalent to imposing   
that the vector $M$ really acts as a true lens: We require big arcs   
with large magnifications and multiple images separated   
by arcsec or even arc-min to {\it focus} into a rather compact   
region in the source plane. This minimization process assumes that   
we are able to associate the multiple lensed images with a particular   
source. This can be achieved with either spectroscopy or with 
morphology and the multicolor imaging of the arcs.  
  
To minimize the variance in the source plane it is illustrative to   
follow the steepest descent path although other more effective minimization   
algorithms can be used (see below).   
Given an initial guess for the mass vector, one can calculate the derivative   
of the variance in the source plane as a function of the mass and minimize   
in the direction of the derivative.   
Once a minimum is found, we calculate the derivative in the   
new mass position and minimize again in an iterative procedure.  
  
The quantity to be minimized is :  
\begin{equation}  
f(M) = \sum_s \sigma _s^2  
\end{equation}  
where the sum is over the number of identified sources and $\sigma _s^2$   
is the variance of the source $s$ in the source plane. That is;  
\begin{equation}  
\sigma _s^2 = < \beta ^2 >_s - < \beta >_s^2  
\label{eq_var_b}  
\end{equation}  
where the $\beta$'s are calculated from equation \ref{eq_lens_matrix} and   
the average is over the $\beta$'s corresponding to source $s$.  
By combining equations \ref{eq_lens_matrix} and \ref{eq_var_b} is easy   
to compute the derivative of $\sigma _s^2$ with respect to $M$.  
\begin{equation}  
\frac{\partial \sigma _s^2}{\partial M_j} = 2 < \beta >_s < \Upsilon _j >_s   
- 2 < \beta \Upsilon_j >_s  
\label{eq_derivM}  
\end{equation}  
where $\Upsilon_j$ is the column $j$ of the $\Upsilon$ matrix and the average   
is made only on the elements associated with the source $s$.   
We should note that all equations involving the vectors $\beta$, $\theta$ or   
$\alpha = \Upsilon M$ have two components, $x$ and $y$ so there will be in fact   
two equations like equation \ref{eq_derivM}. One for the $x$ component of $\beta$ and the   
other one for the $y$ component. At the end, the quantity we aim to minimize is   
$\sigma^2 = \sigma_x^2 + \sigma_y^2$.   
As we already mentioned, the minimization can be done following the path of   
steepest descent given by equation \ref{eq_derivM}. This path will   
end in a minimum at the new mass $M^j$. The process can be repeated by evaluating   
the new path at the new mass position until the variance is   
smaller than certain $\epsilon$. A good choice for $\epsilon$ is to take a few times   
the expected variance for a population of $N_s$ galaxies at the measured $N_s$ redshifts.  
specific values for $\epsilon$ will be discussed later.  
\begin{figure}  
   \epsfysize=8.cm   
%   \begin{minipage}{\epsfysize}\epsffile{Mass_MinVar_Pavlos.ps}\end{minipage}  
   \begin{minipage}{\epsfysize}\epsffile{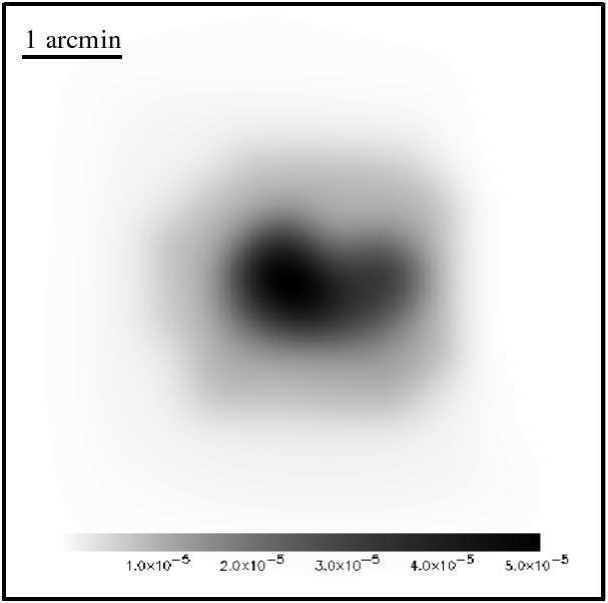}\end{minipage}  
   \caption{  
            Smooth version of the recovered mass after minimizing the variance.  
            The total recovered mass is $1.01 \times 10^{15}$   
            h$^{-1}$ M$_{\odot}$. Compare this mass with the original   
            one in figure \ref{fig_mass}.   
           }  
   \label{fig_recovM_minVar}  
\end{figure}  
\begin{figure}  
   \epsfysize=8.cm   
%   \begin{minipage}{\epsfysize}\epsffile{Source_MinVar_Pavlos.ps}\end{minipage}  
   \begin{minipage}{\epsfysize}\epsffile{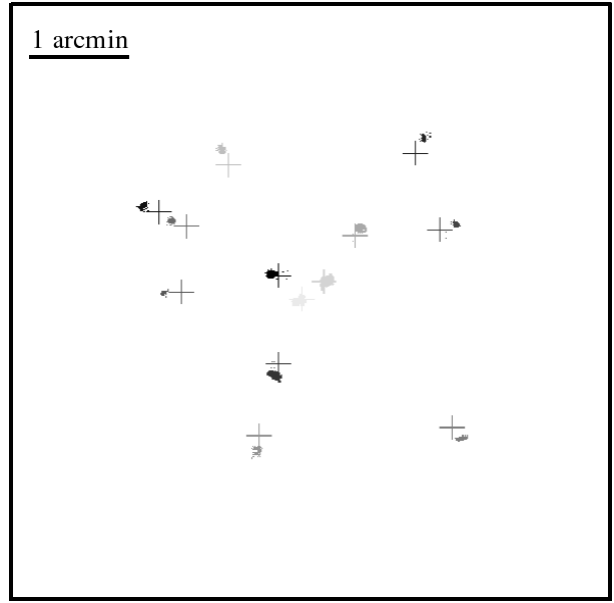}\end{minipage}  
   \caption{  
            Recovered sources after minimizing the variance.  
	    The real positions of the sources are shown as crosses.  
	    Note that the recovered sources are farther away   
            than the real ones. This compensates for the fact that the 
            recovered mass is lower than the the true value.  
           }  
   \label{fig_recovM_minVar}  
\end{figure}  
In practical terms the minimization is done through a series of 
iterations, gradually improving the dynamical grid as laid out in 
section \ref{sect_grid}. For each iteration the ability of the mass 
distribution to focus the $\beta$s into compact sources is improved. 
 
Equation \ref{eq_var_b} can be improved by weighting it 
with the amplification of the different images, assuming the
errors in the image plane are similar from images to images. This point 
will not be explored here.  
The minimization of the variance is a powerful and robust method
for finding a first guess for the mass vector without making 
assumptions about the sources.    
In fact, since the source positions can be estimated from   
equation \ref{eq_lens_matrix}, the minimization of the variance   
also provides us with an initial guess for these.  The drawback is the 
slow convergence of the algorithm.  A typical minimization may take 
several hours on a 1 GHz processor.   
In the next sub-section we will go a step further and we will   
include the $\beta$ positions in the minimization as well as speed   
up the convergence by orders of magnitude.   
  
%---------------------------------  
\subsection{Biconjugate Gradient}\label{sect_conj}  
%---------------------------------  
Inversion of linear systems where the matrix dimensions are   
of order $10^3$, is a numerically trivial problem for today's   
computers provided the matrix is well behaved. If the matrix   
has null or negative eigenvalues, a direct inversion is not   
feasible and one has to aim to solve for some approximated   
solution. Our system of linear equations is a good example   
of an ill-conditioned one. Direct inversion of the   
matrix is not possible due to negative eigenvalues. However there   
is another important reason why we do not want to solve exactly   
(or invert) the system of equations. An exact solution means that   
we will recover a mass distribution which puts the arcs into delta   
function sources. As we will see later, this solution will be   
unphysical. Instead, we are interested in an approximated   
solution which does not solve exactly the system of equations and   
which has a residual. This residual will have the physical meaning   
of the extension of the sources or the difference between the   
point-like sources and the real, extended ones.   
The biconjugate gradient (Press et al. 1997) will be a useful way to {\it regularize}   
our problem. 
 
The biconjugate gradient algorithm is one of the fastest and most   
powerful algorithms to solve for systems of linear   
equations. It is also extremely useful for finding approximate   
solutions for systems where no exact solutions exist or   
where the exact solution is not the one we are interested in.   
The latter will be our case. Given a system of linear equations;  
\begin{equation}  
Ax = b  
\label{eq_conj}  
\end{equation}  
a solution of this system can be found by minimizing the following   
function,  
\begin{equation}  
f(x) = c - bx + \frac{1}{2}x^tAx   
\label{eq_fx}  
\end{equation}  
where $c$ is a constant.  
When the function $f(x)$ is minimized, its gradient is zero.  
\begin{equation}  
\nabla f(x) = Ax - b = 0  
\end{equation}  
That is, at the position of the minimum of the function $f(x)$  
we find the solution of equation (\ref{eq_conj}). In most cases,   
finding the minimum of equation \ref{eq_fx} is much easier than   
finding the solution of the system in \ref{eq_conj} especially when no exact   
solution exists for \ref{eq_conj} or $A$ does not have an inverse. 
  
The biconjugate gradient finds the minimum of equation \ref{eq_fx}   
(or equivalently, the solution of equation \ref{eq_conj}) by   
following an iterative process which minimizes the function   
$f(x)$ in a series of steps no longer than the dimension of   
the problem. The beauty of the algorithm is that the successive   
minimizations are carried out on a series of orthogonal conjugate   
directions , $p_k$, with respect to the metric $A$. That is,  
\begin{equation}  
p_iAp_j = 0 \ \ \ j<i  
\end{equation}  
This condition is useful when minimizing in a multidimensional   
space since it guarantees that successive minimizations do not spoil the   
minimizations in previous steps. 

Let us now turn to the system we want to solve, namely equation \ref{eq_lens2}.  
The biconjugate gradient method assumes that the matrix $\Gamma$ (
matrix $A$ in equation \ref{eq_conj}) is 
square. 
For our case this does not hold since we typically have $N_{\theta} >> (N_c + N_s)$.         
Instead we build a new quantity, called the square of   
the residual,$R^2$:   
\begin{eqnarray}  
R^2 & = &(\theta - \Gamma X)^T(\theta - \Gamma X)  \\ %%%\nonumber  
    & = & 2( \frac{1}{2}\theta^T\theta -\Gamma^T\theta X + \frac{1}{2}X^T\Gamma^T\Gamma X)  
\label{eq_R2}  
\end{eqnarray}  
By comparing equations \ref{eq_R2} and \ref{eq_fx} is easy to identify
the terms,   
$c = \frac{1}{2}\theta^T\theta$, $b = \Gamma^T\theta$ and $A = \Gamma^T\Gamma$.   
Minimizing the quantity $R^2$ is equivalent to solving %solution of equation   
equation \ref{eq_lens2}. %which is also solution of equation \ref{eq_conj}.  
To see this we only have to realize that
\begin{equation}  
b - AX = \Gamma^T(\theta - \Gamma X) = \Gamma^TR  
\end{equation}  
\begin{figure}  
   \epsfysize=8.cm   
%   \begin{minipage}{\epsfysize}\epsffile{Mass_BiConjGrad.ps}\end{minipage}  
   \begin{minipage}{\epsfysize}\epsffile{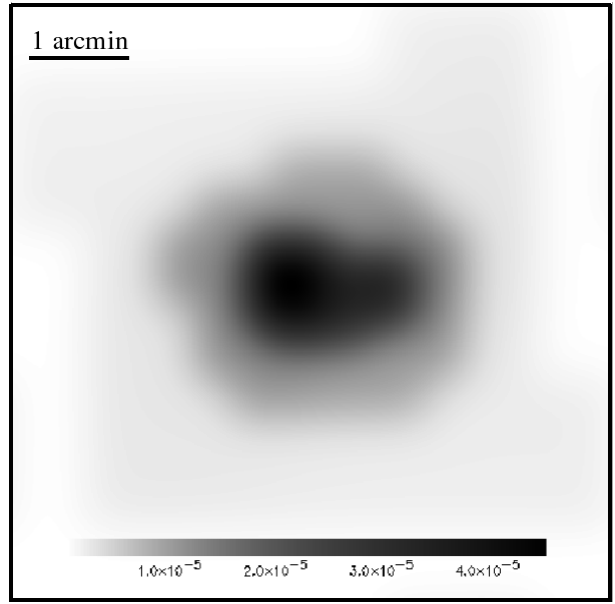}\end{minipage}  
   \caption{  
            Recovered mass after minimizing $R^2$ using the biconjugate gradient   
            algorithm. The mass has been smoothed with a Gaussian filter.  
            The total recovered mass is $1.17 \times 10^{15}$   
            h$^{-1}$ M$_{\odot}$.  
           }  
   \label{fig_recovM_ConjGrad}  
\end{figure}  
\begin{figure}  
   \epsfysize=8.cm   
%   \begin{minipage}{\epsfysize}\epsffile{Source_BiConjGrad.ps}\end{minipage}  
   \begin{minipage}{\epsfysize}\epsffile{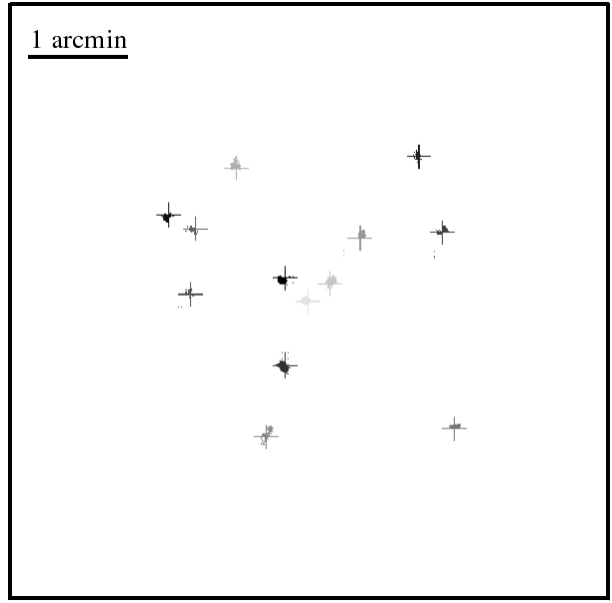}\end{minipage}  
   \caption{  
            Recovered $\beta$'s after minimizing $R^2$.  
	    Again, crosses represent the true position of the sources.   
	    The recovered $\beta_o$ falls in the middle   
            of its corresponding cloud $\beta$ points.  
           }  
   \label{fig_recovB_ConjGrad}  
\end{figure}  
If an exact solution for equation \ref{eq_lens2} does not exist,   
the minimum of $R^2$ will give the better approximated solution to the system.   
The minimum can be now found easily with the biconjugate gradient (Press et al. 1997).   
For the case of symmetric matrices $A$, the algorithm constructs two sequences   
of vectors $r_k$ and $p_k$ and two constants, $\alpha_k$ and $\beta_k$;  
\begin{equation}  
\alpha_k = \frac{r_k^Tr_k}{p_k^TAp_k}  
\end{equation}  
\begin{equation}  
r_{k+1} = r_k - \alpha_k A p_k  
\end{equation}  
\begin{equation}  
\beta_k = \frac{r_{k+1}^Tr_{k+1}}{r_k^Tr_k}  
\end{equation}  
\begin{equation}  
p_{k+1} = r_{k+1} + \beta_kp_k  
\end{equation}  
At every iteration, an improved estimate of the solution is found by;  
\begin{equation}  
X_{k+1} = X_k + \alpha_kp_k  
\end{equation}  
The algorithm starts with an initial guess for the solution, $X_1$, and   
chooses the residual and search direction in the first iteration to be;  
\begin{equation}  
r_1 = p_1 = b - AX_1  
\end{equation}  
Note that $p_1$ is nothing but $ \nabla R^2$.   
Thus the algorithm chooses as a first minimization direction the gradient   
of the function to be minimized at the position of the first guess. Then   
it minimizes in directions which are conjugate to the previous ones until   
it reaches the minimum or the square of the residual $R^2$ is   
smaller than certain value $\epsilon$. 
 
\begin{figure}  
   \epsfysize=8.cm   
%   \begin{minipage}{\epsfysize}\epsffile{Mass_ConjGrad_PointSourceSol.ps}\end{minipage}  
   \begin{minipage}{\epsfysize}\epsffile{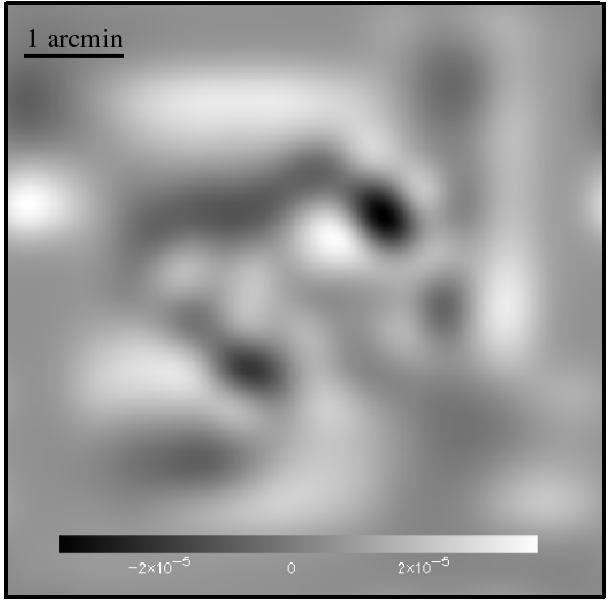}\end{minipage}  
   \caption{  
            Recovered mass after minimizing $R^2$ using the biconjugate gradient   
            algorithm and with a very small $\epsilon$ ({\it point source solution}).  
            The total recovered mass is $2.43 \times 10^{15}$   
            h$^{-1}$ M$_{\odot}$ but there are also regions with negative masses.  
           }  
   \label{fig_recovM_ConjGrad_PS}  
\end{figure}

\begin{figure}  
   \epsfysize=8.cm   
%   \begin{minipage}{\epsfysize}\epsffile{Source_ConjGrad_PointSourceSol.ps}\end{minipage}  
   \begin{minipage}{\epsfysize}\epsffile{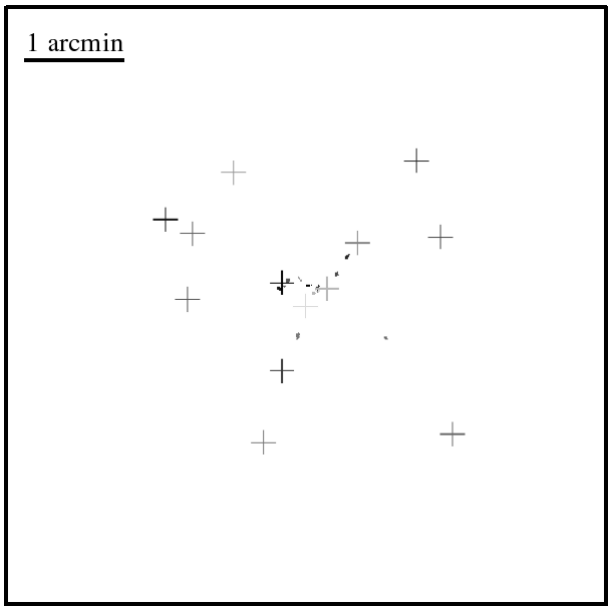}\end{minipage}  
   \caption{  
            Recovered $\beta$'s after minimizing $R^2$ for the {\it point source solution}.  
	    The real source positions are shown as crosses.  
           }  
   \label{fig_recovB_ConjGrad_PS}  
\end{figure}

The method has one potential pathological behavior when applied to our 
problem. One can not choose $\epsilon$ to   
be arbitrarily small. If one chooses a very small $\epsilon$ the   
algorithm will try to find a solution which focuses the arcs in $N_s$   
sources which are delta functions. This is not surprising as we are assuming   
that all the $2N_{\theta}$ unknown $\beta$s are reduced to just $2N_s$ $\beta$s,      
i.e the {\it point source solution} (see figures \ref{fig_recovM_ConjGrad_PS} 
and \ref{fig_recovB_ConjGrad_PS} ) .   
The mass distribution which accomplishes this is usually very much biased compared   
to the right one. It shows a lot of substructure and it has large fluctuations   
in the lens plane. One therefore has to choose $\epsilon$ with   
a wise criteria. Since the algorithm will stop when $R^2 < \epsilon$ we should   
choose $\epsilon$ to be an estimate of the expected dispersion of the sources   
at the specified redshifts. This is the only prior which has to be given to the   
method. However, we will see later how the specific value of $\epsilon$ is not   
very critical as long as it is within a factor of a few of the right source dispersion 
(see figure \ref{fig_Mthr} below).   
Instead of defining $\epsilon$ in terms of $R^2$ is better to define it   
in terms of the residual of the conjugate gradient algorithm, $r_k^2$. This will speed   
the minimization process significantly since we do not need to calculate the real dispersion   
at each step but to use the already estimated $r_k$.   
Both residuals are connected by the relation,  
\begin{equation}  
r_k = \Gamma^TR  
\label{eq_rk}  
\end{equation}  
Imposing a prior on the sizes of the sources means that we expect the residual   
of the lens equation, $R$, to take typical values of the order of the expected dispersion   
of the sources at the measured redshifts. Hence we can define an $R_{prior}$ of the form;  
\begin{equation}  
R_{prior}^i = \sigma_i*RND  
\end{equation}  
where the index $i$ runs from 1 to $N_{\theta}$ and $\sigma_i$ is the  
dispersion (prior) assumed    
for the source associated to pixel $i$ and $RND$ is a random number  
uniformly distributed over -1 and 1.    
Then, we can estimate $\epsilon$ as;  
\begin{equation}  
\epsilon = r_k^Tr_k = R^T_{prior}\Gamma \Gamma^TR_{prior}  
\end{equation}  
When calculating $\epsilon$ is is better to consider only   
the first $N_c$ columns of $\Gamma$ and discard the last $2N_s$. This is recommended   
to avoid that the 1s in this part of the $\Gamma$ matrix do not cancel out when   
multiplied by the random number and dominate the much smaller $\Gamma_{ij}$ elements   
corresponding to the mass components. The last $N_c$ columns of $\Gamma$ should give   
0 contribution when multiplied by the random $R_{prior}^i$ vector.   
If we chose as a prior that the sources are Gaussians with a $\sigma = 30 h^{-1}$ kpc located   
at the measured redshift, this renders $\epsilon \approx 2\times 10^{-10}$.   
The reader will note that the chosen $\sigma$ is a few times larger 
than the one we would assign   
to a typical galaxy. We will discuss this point later.   
The final result is shown in figures \ref{fig_recovM_ConjGrad}  and 
\ref{fig_recovB_ConjGrad}.

One has to be careful in not choosing the $\sigma_i$ very small. In fact, they should   
be larger than the real dispersion of the source. Only when the number of grid   
points, $N_c$, is large enough, can the gridded version of the true mass focus the arcs   
into sources which are similar in size to the real ones. If $N_c$ is not large enough,   
the gridded version of the true mass focuses the arcs into sources which are larger than the   
real sources. We should take this into account when fixing $\sigma_i$.   
  
The reader can argue that a more clever way of including this prior   
information in the algorithm is by perturbing the $\beta$ elements in equation   
\ref{eq_lens2} (or similarly, equation \ref{eq_lens2} in Appendix A). 
This is done by adding some noise to the 1s in the two $\hat{1}$   
matrices in equation \ref{eq_lens3} (see Appendix A). One could for instance add Gaussian noise   
with a dispersion similar to the expected dispersion of the source at redshift $z$.   
The reality however is that the quadratic nature of $R^2$ cancels out any   
symmetric perturbation added to the elements of $\Gamma$. Thus, the result   
is similar if we perturb $\Gamma$ or not and we still have to include the   
prior and fix $\epsilon$ to be large enough so we do not recover   
the {\it point source solution}. This also tell us that this method is not   
very promising if one wants to include parity information in the recovery of the   
mass and sources. In the next subsection we will show a different approach   
which can include this parity information.   
%-----------------------------------------  
\subsection{Singular Value Decomposition} \label{svd}  
%-----------------------------------------  
The Singular Value Decomposition  (hereafter SVD) 
algorithm  allows for decomposition of a generic $m\times n$   
(with $m >= n$) matrix $A$ into the product of 3 matrices, two orthogonal and   
one diagonal (e.g Press et al. 1997).  
\begin{equation}   
A = U W V^T  
\end{equation}  
where $U$ is an $m\times n$ orthogonal matrix, $W$ is an $n\times n$   
diagonal matrix whose elements in the diagonal are the {\it singular   
values} and $V^T$ is the transpose of an $n\times n$ orthogonal matrix.  
When $A$ is symmetric, the SVD reduces to finding the {\it eigenvectors}    
and {\it eigenvalues} of A. 
 
The advantage of this decomposition is that the inverse of $A$ is   
given by ;  
\begin{equation}   
A^{-1} = V W^{-1} U^T  
\label{eq_SVD_Inv}  
\end{equation}  
where $W^{-1}$ is another diagonal matrix whose elements are just   
the inverse of the elements of $W$, that is $W^{-1}_{jj} = 1.0/W_{jj}$.   
The proof $A^{-1}A=I$ follows from the property that $U$ and $V^T$ are   
orthogonal ($U^T U = V V^T = I$). 

In our case, we can use SVD to calculate the inverse of the $\Gamma$ matrix and 
find the solution, $X$, directly by inverting equation (\ref{eq_lens2})
\begin{equation}
X = \Gamma^{-1} \theta = V W^{-1}U^T \theta
\end{equation} 
Although the SVD allows us to invert the problem by calculating   
$\Gamma^{-1}$, its full power lies in its ability to solve a system   
approximately. The level of approximation can be controlled by setting a   
threshold in the matrix $W^{-1}$. In our problem, there will be many   
equations which are strongly correlated in the sense that most of the   
$\theta$ positions in a single arc will come from the same source (that   
is, they will have almost the same $\beta$). Also we have to keep in   
mind that we are using all the pixels in our data. This means that two   
equations corresponding to two neighboring pixels will look almost   
exactly the same. When one computes the SVD of the matrix $\Gamma$, these   
two facts translate into a matrix $W$ with elements in the diagonal which are   
0 or close to 0. The inverse of $W$ would be dominated by these small   
numbers and the solution will look very noisy. The good news about using   
SVD is that the most relevant information in the $\Gamma$ matrix is packed   
into the first few values (the largest values in $W$) while the small   
single values in $W$ will contain little information or the redundant   
information coming from neighboring pixels. One can just approximate $W$   
by another matrix $W'$ where all the elements in its diagonal smaller   
than certain threshold are set to 0. Also, in the inverse of $W'$, these   
elements are set to 0. The magic of this trick is that the solution found   
with this approximation will contain the main trend or main components of the   
mass distribution. 
 
\begin{figure}  
   \epsfysize=8.cm   
%   \begin{minipage}{\epsfysize}\epsffile{Mass_SVD_NoPrior.ps}\end{minipage}  
   \begin{minipage}{\epsfysize}\epsffile{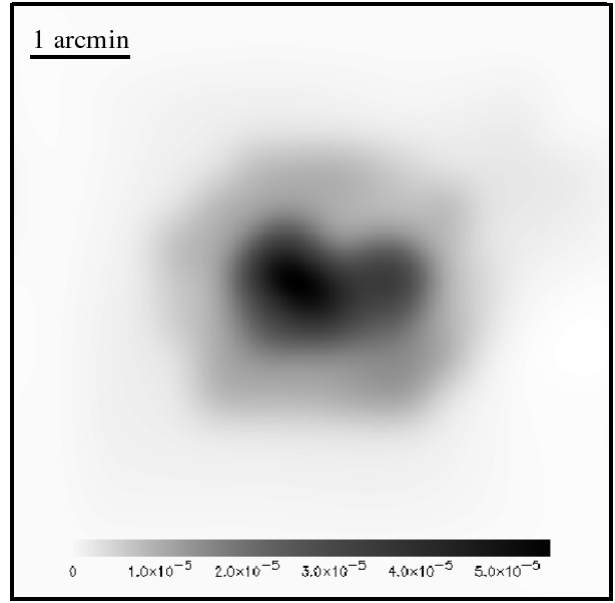}\end{minipage}  
   \caption{  
            Recovered mass after SVD (no prior).  
            The total recovered mass is $1.01 \times 10^{15}$   
            h$^{-1}$ M$_{\odot}$.  
           }  
   \label{fig_recovM_SVD}  
\end{figure}

\begin{figure}  
   \epsfysize=8.cm   
%   \begin{minipage}{\epsfysize}\epsffile{Source_SVD_NoPrior.ps}\end{minipage}  
   \begin{minipage}{\epsfysize}\epsffile{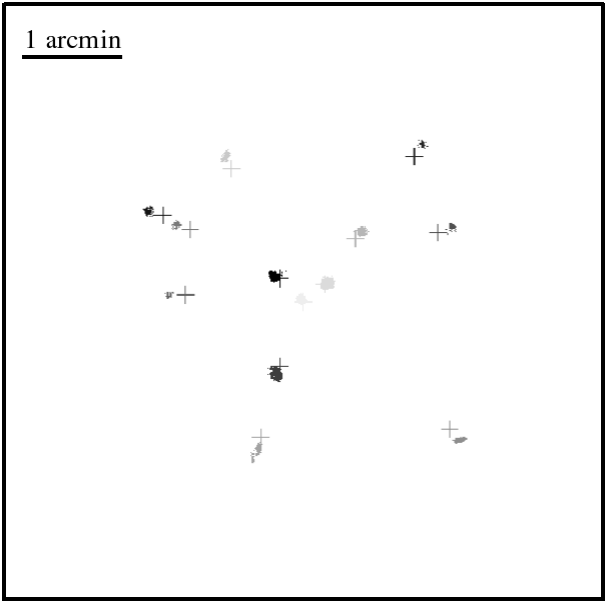}\end{minipage}  
   \caption{  
            Recovered sources after SVD (no prior).  
	    The real positions of the sources are shown as crosses  
           }  
   \label{fig_recovB_SVD}  
\end{figure}

Another advantage of using the SVD algorithm is that in this case no   
prior on the extension of the sources is needed. The degree of accuracy is controlled by   
setting the threshold in the singular values of the matrix $W$. Those elements in $W$ below   
the threshold are set to 0 and the same in its inverse. The threshold is usually set after   
looking at the singular values. The first ones will normally stay in some kind of   
{\it plateau} and after then the singular values will decrease rapidly. The threshold   
should be normally set immediately after the {\it plateau}. In figures   
\ref{fig_recovM_SVD} and \ref{fig_recovB_SVD} we show the result after decomposing $\Gamma$   
in its SVD decomposition and calculating its inverse with (\ref{eq_SVD_Inv}). 
  
Like the two previous algorithms, the SVD has its own pathologies. Using standard   
subroutines to find the SVD of $\Gamma$ usually return a {\it no convergence} error.   
This error comes from the nearly degenerate nature of $\Gamma$. One has then to increase   
the number of maximum iterations on these subroutines or use a coarse version of   
$\Gamma$ where only a small fraction of the $\theta$ positions (or  
equivalently, $\Gamma$ rows)    
are considered. Another solution is inverting the preconditioned 
system of equations where    
we previously multiply for $\Gamma^T$. This allows to use all the $\theta$ pixels and find   
the SVD of $\Gamma^T\Gamma$ in a small number of iterations. However,  
using the SVD of $\Gamma$    
instead of $\Gamma^T\Gamma$ has a very interesting feature. It allows  
to introduce parity information    
in an effective way. 
 
Contrary to the quadratic cases of minimization of the variance or   
the square of the residual $R^2$, the SVD allows us to include parity information   
in the $\Gamma$ matrix which will not disappear when we look for the best solution.   
Since no $\Gamma^T\theta$ or $\Gamma^T\Gamma$ operations are involved, parity   
information will not cancel out. SVD could be an interesting way of   
fine tuning the solution by including the extra information coming from the   
parity of the arcs.   
  
%%%%%%%%%%%%%%%%%%%%%%%%%%%%%%%%%%%%%%%  
\section{Incorporating the null space} \label{nullspace}  
\begin{figure}  
   \epsfysize=8.cm   
%   \begin{minipage}{\epsfysize}\epsffile{RecoveredArcs.ps}\end{minipage}  
   \begin{minipage}{\epsfysize}\epsffile{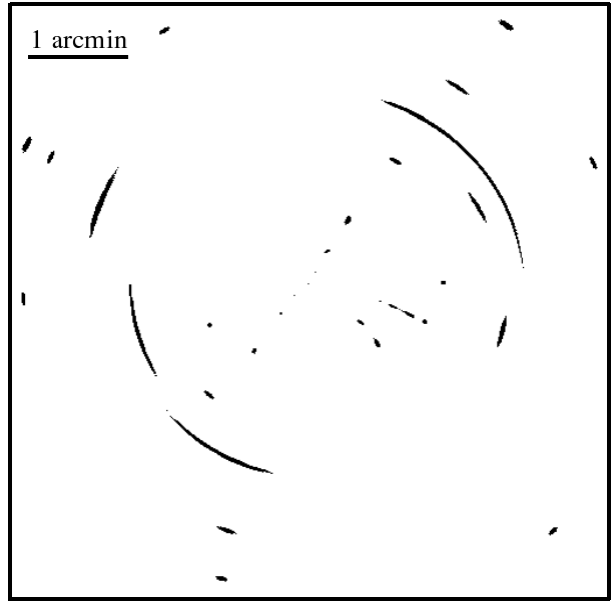}\end{minipage}  
   \caption{The plot shows the predicted arcs according to the best  
     mass and source solution. We have assumed that the sources are  
     circles with radius $15 h^{-1}$kpc centred in the source position  
     and at the measured redshifts. This result should be compared  
     with the true arcs seen in figure \ref{fig_theta}. The match is 
     almost perfect.  
           }  
   \label{fig_recoveredArcs}  
\end{figure}

%%%%%%%%%%%%%%%%%%%%%%%%%%%%%%%%%%%%%%%  
So far we have made use of the information contained in the observed strong   
lensing arcs. This gives us a solution which explains the data in the sense   
that it predicts arcs in the right positions but it could well happen that   
the solution over-predicts arcs. An example of this can bee seen by comparing  
the reproduced arcs in fig. (\ref{fig_recoveredArcs}) with the true  
arcs in fig. (\ref{fig_theta}). Some of the reproduced arcs are  
slightly larger and span a larger area than the true ones,  
although the sources are perfectly reproduced in their true positions  
and the recovered mass is  
very close to the true mass distribution.  
  
To avoid this we propose for the first time to include information   
from the null space, $\tilde{\theta}$. That is, the part of the image 
which does not contain any arc.  This space tell us where an arc should   
not appear but it does not tell us where the hypothetical $\tilde{\beta}$   
should be. The only fair thing we can do is to impose that none of the pixels   
in the null space fall into the estimated $\beta$ of our solution. By achieving   
this we will have a solution which predicts the right arcs while not 
over-predicting arcs. The solution will be then fully consistent with 
the observed data.   
  
The null space is connected to the solution X by;  
\begin{equation}  
\tilde{\beta} = \tilde{\theta} - \tilde{\Upsilon}M  
\label{eq_tilde}  
\end{equation}  
It is evident that we want the new solution X ($M$ and $\beta_o$) to be such that    
the new $\tilde{\beta}$ do not fall within a circle of radius $p(k)$  
centered in each one of the $\beta_o$ where $p(k)$ is the prior with  
information on how extended are the sources.  
The null space will perturb the solution X in such a way that the new solution,   
$X' = X + \Delta \tilde{X}$ is an approximated solution of equation \ref{eq_lens2}   
and satisfies all the constraints of the form ${\tilde{\beta}} \ni {\beta}$.  
  
The way we incorporate the constraints is by adopting an approach commonly used in      
penalized quadratic programming. We will minimize the new function  
\begin{equation}  
\phi(X) = R^2 + \lambda \sum_k g(f_k)   
\label{eq_phi}  
\end{equation}  
where the $\lambda$ is a constant which guarantees that in the first iterations,   
the second term in equation \ref{eq_phi} does not dominate the first and   
$g_k$ is a function which will penalize those models   
predicting $\tilde{\beta}$ falling near the $\beta$ positions which minimize   
$R^2$. As a penalizing function we will choose an asymptotically divergent   
Gaussian  
\begin{equation}  
g(f_k) = \frac{1}{e^{f_k} - \mu}  
\end{equation}  
with,  
\begin{equation}  
f_k = \frac{(\tilde{\beta}(k)_x - \beta_x^o)^2 + (\tilde{\beta}(k)_y - \beta_y^o)^2}{\sigma _k^2}   
\end{equation}  
Where $\tilde{\beta}(k)_x$ is the x component of $\tilde{\beta}$ for the pixel $k$   
in the null space. $\beta_x^o$ is our estimated value of $\beta$ (x component) and    
similarly for the y component. There are as many constraints of the form $f_k$   
as there are pixels, $\tilde{\theta}$, in the null space.  
  
\begin{figure}  
   \epsfysize=6.cm   
   \begin{minipage}{\epsfysize}\epsffile{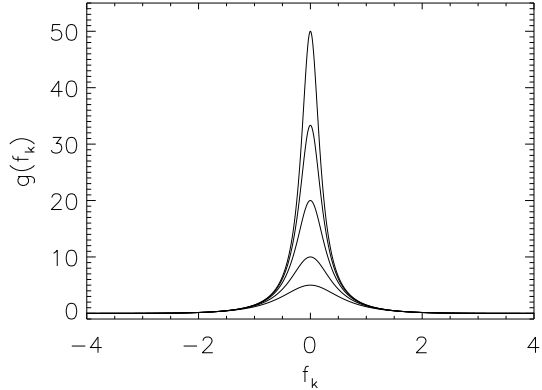}\end{minipage}  
   \caption{  
            Penalizing function showing values of $\mu = 0.8, 0.9, 0.95,   
            0.97,$ and 0.98 and $\sigma = 1$. Note that the width of the   
            curve does not change when increasing $\mu$.   
           }  
   \label{fig_Gauss}  
\end{figure}

The parameter $\mu$ controls the degree of divergence of the Gaussian function.   
When $\mu=0$ we recover the classical Gaussian but as $\mu$ approaches 1 the Gaussian   
becomes more and more sharply without increasing its dispersion.   
For $\mu=1$ the Gaussian is infinite at $f_k = 0$ (see figure \ref{fig_Gauss}.   
By minimizing the function $\phi(X)$   
with increasing values of $\mu$ we will find that in the limit $\mu \rightarrow 1$ the   
solution will push away those $\tilde{\beta}$ which originally were falling in the region   
defined by the set of {$\beta$}.  
  
Ideally we want to include all the {\it dark} or empty pixels in the $\tilde{\theta}$   
space but this is, in general, found to be a waste of memory and computing time.   
The fact is that only the $\tilde{\theta}$ pixels which are hitting (or close to hit)   
one of the $N_s$ estimated $\beta$ positions of the sources will have  
some impact on the solution.    
Most of the $\tilde{\theta}$ pixels in the null space already satisfy all the constraints   
for the actual solution X. For this reason we will include only   
those $\tilde{\theta}$ for which the solution X predicts their  
corresponding $\tilde{\beta}$     
are close to hitting (or actually hitting) a source. The  
$\tilde{\theta}$ space will include the observed    
$\theta$'s as a subspace. We have to exclude this subspace from  
$\tilde{\beta}$  before minimizing  equation \ref{eq_phi}.  This is, again, 
just an optimization process. In practice, one can include all the pixels in the image 
(excluding those containing part of an arc) in the null space.
 
After the minimization process is finished, the new solution will have the arcs falling in   
compact regions around $\beta_0$ while the extra-arcs produced by the previous solution will   
fall in regions outside areas around the $\beta_o$. 
 
>From our simulation we have seen that addition of the null space induces small   
changes in the mass plane and it tends to stabilize the solution in the sense   
that it makes the recovered mass profile independent of the threshold   
$\epsilon$. This is in fact an interesting bonus which comes with the addition of the   
null space. The new function to be minimized, $\phi$, can be minimized until the true   
minimum  is found. In this case, there is no equivalent of a point source solution.   
Since some of the $\tilde{\theta}$ are in fact very close to the observed $\theta$,   
the solution which minimizes $\phi$ will {\it focus} those  $\tilde{\theta}$ and   
$\theta$ in neighboring regions in the source plane. When minimizing $\phi$ there   
will be two competing effects. One will tend to increase the mass so it minimizes   
$R^2$ (point source solution). The other will tend to reduce the mass so the   
$\tilde{\beta}$ will be pushed away from the $\beta$ positions. The outcome will be   
a balanced situation between the $\beta$ trying to collapse into compact sources and the   
$\tilde{\beta}$ trying to escape the wells in the $beta$ positions.   
  
\begin{figure}  
   \epsfysize=8.cm   
%   \begin{minipage}{\epsfysize}\epsffile{RecovM_BetaTilde.ps}\end{minipage}  
   \begin{minipage}{\epsfysize}\epsffile{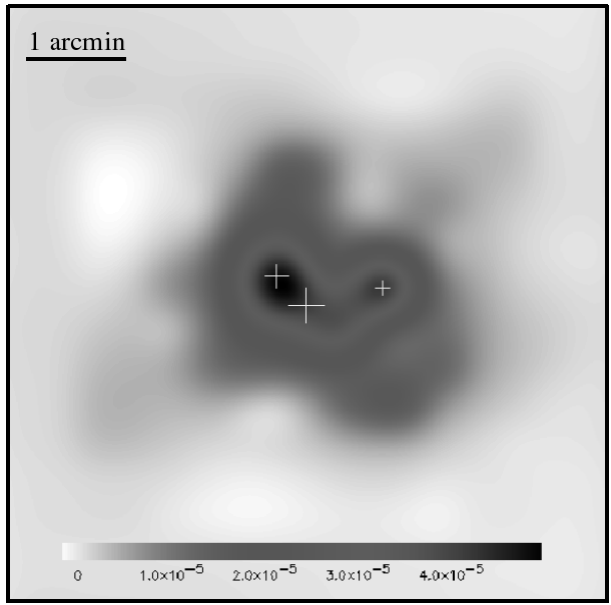}\end{minipage}  
   \caption{  
            Recovered mass (smoothed) after minimizing $\phi$. The crosses show   
            the position of the three main halos in the cluster. The size of the   
            cross is proportional to the mass of the halo. The total recovered   
            mass is $1.1055 \times 10^{15} h^{-1} M_{\odot}$.    
           }  
   \label{fig_RecovMTilde}  
\end{figure}  
  
To accurately quantify the effect more simulations are needed. This will be done   
in a future paper but as an illustration we show in figure \ref{fig_RecovMTilde}   
the recovered mass after imposing the constraints in the $\tilde{\theta}$ space.   
The total mass is now only 1.2 \% lower than the true mass. The new mass also   
contains more structure and starts showing the internal distribution of the main   
components of the cluster. 
 
In figure \ref{fig_recoveredArcs} we show the predicted position for the arcs   
after combining the best mass together with the best estimate for   
the position of the sources. To compute the arcs we have assumed that each source   
is a circle with a radius $15 h^{-1}$kpc centred in the estimated best source   
positions and at the measured redshifts. By comparing figures \ref{fig_recoveredArcs}   
and figure \ref{fig_theta} (left) we see that the predicted arcs matches very well   
the observed (simulated) data with the exception of some of the arcs near the   
center of the image.  
  
%%%%%%%%%%%%%%%%%%%%  
\section{Discussion}  
\begin{figure}  
   \epsfysize=6.cm   
   \begin{minipage}{\epsfysize}\epsffile{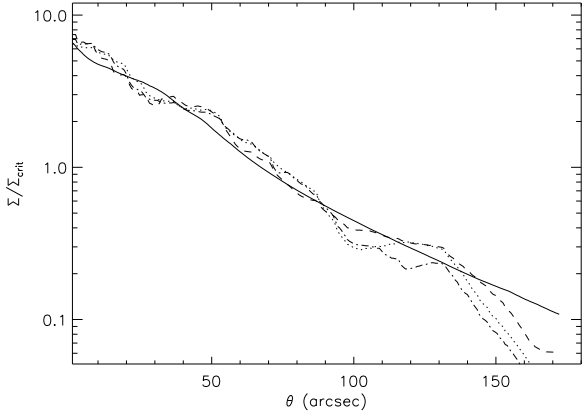}\end{minipage}  
   \caption{  
            Original profile (solid line) compared with the recovered ones. 
            Dashed line is the biconjugate gradient case, dot-dashed the SVD case, 
	    dotted the case is when include the null space. The minimum of the variance 
            is similar but with less mass in the tails.  
            All profiles have been normalized by the critical 
            surface density for a source at redshift, $z_s = 3$.
           }  
   \label{fig_prof}  
\end{figure}

%%%%%%%%%%%%%%%%%%%%  
In this paper we have presented several approaches to  
non-parametric lens modelling. Using no information about luminosity  
distributions whatsoever, these methods perform remarkably well on  
simulated strong lensing data. One of the main conclusions of this   
work is simply that it works; It is possible to recover information   
about the mass distribution without using prior information on the   
same if one has a sufficiently large number of arcs.  
This can be seen in figure \ref{fig_prof} where we show the recovered mass 
profiles compared with the original one. The recovered mass traces the real 
mass distribution up to furthest data point in the simulated image. 
Beyond this point, the recovered profile is insensitive  
to the data. When the minimization stops, the cells in the outer 
regions stay with a mass close to the one they had in the first step of 
the minimization. This point is discussed in more detail in Diego et al. (2004).

We are optimistic about the performance of these methods   
when used on future data, but we would also like to emphasize the   
potential pathologies. We will now discuss the major issues.  
As we have seen in section (\ref{linear}) the inversion algorithms  
rely on a linearization of the lens equation. We achieve that   
by decomposing the lens plane into small mass-cells and  
assuming that the gridded mass is a fair representation of the true underlying mass. This is   
true when the number of cells in the grid is large enough, but for a  
uniform grid this may mean using several thousand mass-cells, making  
the problem ill conditioned and underdetermined. By inverting the lens equation   
in a series of iterations we introduce an adaptive grid which optimizes the number of   
cells by sampling dense regions more heavily and using larger cells for  
underdense regions. This allows for good sampling of the lens without  
a huge number of cells.  
However, the gridded mass plane is still an approximation to the true  
mass plane so we expect the solution to be an approximation to the  
true solution as well. Since a solution comprises not only the masses  
but the positions and extents of the sources as well, this means that we should  
expect these to also be approximations to the true ones. 
This has  already been identified as one of the potential   
pathologies of the algorithm. Namely, if we try to focus the sources into very compact   
regions, with sizes comparable to the typical galaxy sizes at the  
relevant redshifts, the obtained mass is in general different to the true mass. 
In fact the best results are obtained when the mass plane focuses the arcs into regions 
which are a few times larger than the extent of the true sources. 
As we pointed out before, this reconstructed mass corresponds to a 
{\it short-sighted} version of the lens.   
This problem can be overcome by requiring the minimization algorithms  
stop once the recovered sources are a few times larger than the true  
sources. The extent of the true sources can be guessed from the redshift.  
\begin{figure}  
   \epsfysize=6.cm   
   \begin{minipage}{\epsfysize}\epsffile{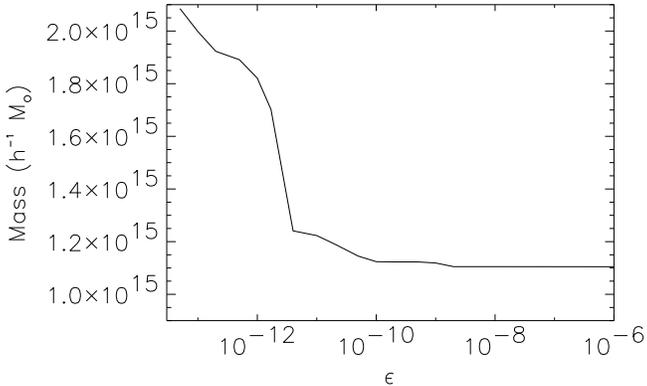}\end{minipage}  
   \caption{  
            Total mass as a function of the threshold $\epsilon$ used in the   
            biconjugate gradient minimization. The true mass is   
            $1.119 \times 10^{15} h^{-1} M_{\odot}$.  
           }  
   \label{fig_Mthr}  
\end{figure}  
  
Introducing such a prior begs the question: How sensitive is the  
solution to the specific guess of prior? The answer is that it depends  
on exactly ``how bad'' that choice is. As long as we assume a source size  
significantly larger than the true sizes the actual solution does not change  
much and resembles well the true mass distribution. However, when  
trying to approach the true source size, the solutions changes rapidly  
away from the realistic model. The situation is shown in  
fig. (\ref{fig_Mthr}). As shown in section \ref{sect_conj} a given physical  
size corresponds to a given threshold, $\epsilon$, and as long as this  
threshold is sufficiently large, $\ge 10^{-10}$, (corresponding to a  
physical size of $\ge 30 h^{-1}$ kpc) the total   
mass is well behaved. If we instead demand that the physical size of  
the reconstructed sources be more realistic, say $\sim 12 h^{-1}$ kpc, we get  
a threshold of $\sim 10^{-11}$ for which the mass distribution is  
already starting to diverge away from the true mass. We  
therefore want the recovered sources to be larger than the true  
ones. In other words we want to recover a \emph{short-sighted} cluster!  
In previous sections we have demonstrated non-parametric lens modelling using several  
different algorithms with promising results. However we have  
not yet addressed the question of uniqueness. Is there a unique  
solution, and if not, how different are the possible solutions?  
  
The good news is that minimizing quadratic functions like the variance or   
the square of the residual guarantees that there is only one absolute minimum.   
This absolute minimum corresponds to the {\it point source solution}.   
The bad news is that this is not the solution we are looking for. As  
shown above, trying to focus the sources too much introduces artifacts  
in the mass distribution, so we need to stop the minimization at some  
step before the absolute minimum. In two dimensions it is easy to  
visualize that the quadratic function will have the shape of a valley,  
and stopping the minimization at some point before the actual minimum  
means choosing an ellipse on which all solutions are equally good. In  
many dimensions this is harder to visualize but we expect our obtainable  
solutions to lie on an N-dimensional ellipsoid around the minimum.  
   
To get a quantitative grasp on how much these solutions differ in mass  
and source positions, we solve the equations for a range of random initial  
conditions. This is a manageable task since our minimization algorithm is  
extremely fast, taking only about $\sim 1$ second on a 1GHz processor. 
Also for speed purposes, we fix the grid to the one corresponding to the solution 
shown in figure  \ref{fig_recovM_ConjGrad}. 
Fixing the grid speeds the process significantly since the $\Gamma^T\Gamma$ matrix and 
the $\Gamma^T\theta$ vector do not need to be recalculated in each minimization. 
In Diego et al. (2004) a similar process is followed carrying multiple minimizations  
but this time the grid is changed dynamically based on the solution found in the previous 
step. The authors found on that paper that the dispersion of the solutions increases due 
to the extra variability of the grid.
 
\begin{figure}  
   \epsfysize=6.cm   
   \begin{minipage}{\epsfysize}\epsffile{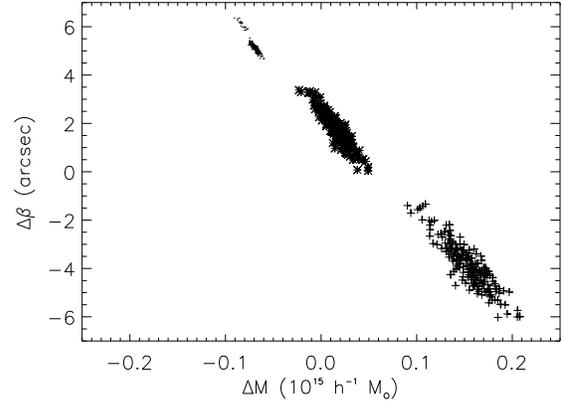}\end{minipage}  
   \caption{  
           Dispersion in the solutions for three different sets of   
           initial  conditions. Upper left (dots), we start with a random realization   
           of small masses. Asterisks in the center, starting from a random realization   
           but with a total mass of $M \approx 1.06$. Crosses (bottom right),   
           starting with high random masses. In all cases, the threshold was set   
           to $\epsilon = 2 \times 10^{-10}$ and the starting $\beta$ positions where   
           chosen as random in a box of $100 \times 100$ centered in the center of the image.  
           }  
   \label{fig_Cloud}  
\end{figure}

The result is shown in figure \ref{fig_Cloud}. The minimization  
process described in section \ref{sect_conj} will stop at a different  
point in the N-dimensional ellipsoid for each set of different initial  
conditions. If the total mass of the initial mass distribution is very  
low the minimization will stop at solutions with mass below the true  
mass and $\beta$ positions further away from the center of the cluster  
than the real ones (dots). If instead we start with a total mass much  
larger than the true value, the minimization process returns higher  
masses and $\beta$ positions closer to the center of the potential  
(crosses). The situation improves when we impose that the total mass  
of the initial distribution have a reasonable value.  
In this case, the minimization stops in a region close to both the  
right mass and the right $\beta$s. This argument motivates an  
iterative minimization where successive estimations for the mass are  
obtained at each step and used in the next. 
  
Among the three algorithms presented in this work, the minimization of   
the variance is the less powerfull due to its low convergence.   
It is however interesting from a pedagogical point of view since   
it shows the existing degeneracies between the total mass of the   
cluster and the positions of the sources. The biconjugate gradient   
is orders of magnitude faster and is capable of finding the point   
source solution in a few seconds. The point source solution is   
however unphysical and a prior associated to the size of the sources   
is needed in order to stop the minimization process at the proper place.   
The good news are that the algorithm shows a weak sensitivity to this   
prior provided it is chosen with a minimum of wisdom. Since the    
point where the minimization stops depends on where the minimization   
starts, this also allows to study the range of possible models consistent   
with the data by minimizing many times while changing the initial   
conditions. This feature makes the biconjugate gradient very attractive   
for studying the space of solutions. This space can be reduced by   
including the extra-information contained in the null space, that is,   
the areas in the sky with no observed arcs. We have seen how   
this information can be naturally included in the minimization   
process by introducing a penalty function. 
 
Finally, the SVD has the interesting feature that it allows to add   
extra information regarding the parity or resolved features in the   
arcs. This possibility has not been studied in detail in this paper   
but is definitely worth exploring since it would allow to recover   
smaller details in the mass distribution. The main drawbacks of   
the SVD is that the decomposition fails to converge when all the   
information is used and one has to use a {\it coarse} version of the   
data instead. The second drawback is that some of the singular values   
in the matrix $W$ are very small. These values will dominate the   
inverse if they are not {\it masked} by a threshold in the matrix   
$W$. Choosing the value for this threshold is not very   
critical as long as its amplitude is large enough to {\it mask}   
the small singular values in $W$.

The accuracy of the methods presented in this paper allow for  
high precision non-parametric mass reconstructions and direct mapping  
of the dark matter distribution of clusters. Previous works have suggested   
some discrepancy between mass estimates derived from different data   
(Wu \& Fang 1997). Accuracy, combined with the speed of the  
algorithm opens the door to cosmological studies. Strong lensing  
analyses were predicted to yield interesting cosmological constraints,  
but due to the uncertainties in our understanding of galaxy clusters they have yet  
to live up to these predictions. A fast, per-cent level determination of  
cluster masses from lensing observations, could allow for sufficient  
statistical sampling to provide information about cosmological  
parameters. An interesting follow up to this work would be to  
establish what number of strong lensing systems, and what quality of  
data is necessary in order to make interesting constraints using our  
methods.   

However the methods presented here should not be applied indiscriminately. For  
instance, if the reconstructed mass is systematically biased toward  
recovering less cuspy central regions this reduces the  
possibility of making conclusions about mechanisms for dark matter  
annihilation (Spergel \& Steinhardt 2000,  
Wyithe et al. 2001, Boehm et al. 2004).   
If one is interested in using our methods to discuss this, that bias must be  
quantified. Due to lens resolution issues we anticipate that our  
reconstruction algorithms indeed have a bias in this direction and it  
is likely that other algorithms may suffer from similar problems.  
These and other poetential systematic errors shall be investigated in future 
works.

%%%%%%%%%%%%%%%%%%%%%%%%%  
\subsection{Future work}  
%%%%%%%%%%%%%%%%%%%%%%%%%  
This paper is not intended to be an exhaustive exploration of the  
accuracy of non-parametric mass reconstruction methods but rather to  
help re-ignite debate and competition in the field, 
and demonstrate the feasibility and power of such methods 
in the face of new data. Specific issues such as  
magnitudes of the mass profile bias, source morphologies,   
surface brightness of the arcs, parity, projection effects as well as other   
relevant issues discussed above should be explored in order to improve   
the results. 
Special emphasis should be paid to the effects the grididification of the 
lens plane has on the results. An attempt to study this issue has been made on 
a second paper (Diego et al 2004) where the authors have shown that the grid 
may play an important role. 
In this paper we have presented the results using 
a very specific simulation. Although the algorithm has been tested with different 
simulations with positive results, it would be desirable to make a more detailed 
study of how the result is affected by issues like the structure of the lens, 
its symmetry or the number of sources and accuracy in their redshifts. Such a study 
would be even more interesting if a comparative study between parametric and 
non-parametric methods is done using the same simulations. 
Also interesting is to explore the effects of adding   
constraints in the mass of the type $M_i > 0$ which may help to stabilize   
the solution even in the absence of a prior on the source sizes.   
This can be accomplished by adopting the techniques used in quadratic   
programming. All theses issues, although very interesting, are beyond the scope 
of this paper and will be studied in subsequent papers. 
  
Although not discussed in this paper, a very interesting piece of work would   
be about the potentiality of an accurate strong lensing analysis as a   
cosmological tool (e.g Link \&  Pierce 1998, Yamamoto et al. 2001, 
Golse et al. 2002, Soucail et al. 2004). 
Previous works (Yamamoto et al. 2001, Chiba \& Takahashi 2002, Sereno 2002, 
Dalal et al. 2004) suggest that   
the lensing observables are primarily dependent on the lens model while   
the dependency in the cosmological parameters is minor. Constraining the   
lens model with accuracy can open the window to do cosmology with   
strong lensing images.   
Is easy to imagine that a single image of a cluster with dozens   
of arcs coming from sources at different redshifts will constrain the lens   
model rather accurately and then, it will have something   
to say about the cosmology given the large number of distances   
involved in the analysis.   
All these issues are of great interest and they are intended to be   
studied in subsequent papers.

%%%%%%%%%%%%%%%%%%%%%%%%%%%  
\section{Acknowledgments}  
%%%%%%%%%%%%%%%%%%%%%%%%%%%  
This work was supported by NSF CAREER grant AST-0134999, NASA grant  
NAG5-11099, the David and Lucile Packard Foundation and   
the Cottrell Foundation. The work was also partially supported by
B. Jain through lousy poker playing. We thank D. Rusin and G. Bernstein for  
helpful discussions. We would like also to thank J.P Kneib for helpful 
suggestions.
  
%%%%%%%%%%%%%%%%%%%%%%%%%%%%%%%%%%%%%%%%%%%%%%%%%%%%%%%%%%%%%%%%%%%%%%  

%%%%%%%%%%%%%%%%%%%%%%%%%%%%%%%%%%%%%%%%%%%%%%%%%%%%%%%%%%%%%%%%%%%%%%%  
\large{\bf APPENDIX A: THE $\Upsilon$ AND $\Gamma$ MATRICIES}\\  
\\  
The $\Upsilon$ matrix contains the information of how each mass element  
$j$ affects the $i^{th}$ deflection angle. That is   
\be  
\alpha_i = \Upsilon_{ij} M_j.  
\ee  
Precisely how the matrix is constructed is a matter of  
convenience. Our specific choice is to put both $x$ and $y$ components  
of all the arc pixels into vectors with $2 N_{\theta}$ elements. The  
resulting $\Upsilon$ is then a $2 N_{\theta} \times N_{cells}$ matrix.  
  
In fact the structure of the $\Upsilon$ matrix and the vectors are irrelevant as  
long as they combine to correctly represent the lens equation as   
\be  
\beta_i = \theta_i - \Upsilon_{ij}M_j.  
\ee  
  
Each element in the $\Upsilon$ matrix is computed as follows.  
\begin{equation}  
\Upsilon_x(i,j) = \lambda[1 - exp(-\delta/2\sigma^2)] \frac{\delta _x}{\delta^2}  
\label{eq_gamma}  
\end{equation}  
where   
\begin{equation}  
\lambda = 10^{15}M_{\odot} \frac{4 G}{c^2}\frac{D_{ls}}{D_l D_s}  
\label{eq_lambda}  
\end{equation}  
The index $i$ runs over the $N_{\theta}$ observed $\theta$ pixels,   
index $j$ runs over the $N_c$ elements in the mass vector, $M$.   
The factor $\delta _x$ in equation (\ref{eq_gamma}) is just the difference   
(in radians) between the x position in the arc (or x of pixel $\theta_i$) and   
the x position of the cell $j$ in the mass grid   
($\delta_x = \theta_x(i) - \theta'_x(j)$).   
Similarly we can define $\delta_y = \theta_y(i) - \theta'_y(j)$ and   
$\delta = \sqrt{\delta_x^2 + \delta_y^2}$. Also, for $\Upsilon_y$   
we only have to change $\delta_x$ by  $\delta_y$.   
Since we include the factor $10^{15}$M$_{\odot}$ in $\lambda$ (see equation   
\ref{eq_lambda}), the mass vector $M$ in equation \ref{eq_lens_matrix}   
will be given in $10^{15}$ h$^{-1}$ M$_{\odot}$ units. The $h^{-1}$   
dependency comes from the fact that in $\lambda$ we have the ratio   
$D_{ls}/(D_l D_s)$ which goes as $h$.   
Also we calculate $\Upsilon_x$ and $\Upsilon_y$ separately, the final $\Upsilon$   
matrix entering in equation \ref{eq_lens_matrix} contains both, $\Upsilon_x$   
and $\Upsilon_y$ (the same holds for the vectors $\beta$ and $\theta$).   
One can rearrange the x and y components in any order.\\  
  
The structure of the $\Gamma$ matrix is identical to the matrix   
$\Upsilon$ but with the difference that it has $2N_s$ additional columns   
(the location of the extra-columns is irrelevant as long as   
it is consistent with the location of the $2N_s$ $\beta _o$ unknowns in the   
$X$ vector).   
Is easy to see that each one of these extra columns (with dimension   
$2*N_{\theta}$) corresponds to one of the $N_s$ sources.   
Since the $\Gamma$ matrix has to contain both the x and y component,   
the first/second half of each one of the extra    
columns will be all 0's depending on if it corresponds to the y/x component   
of $B_o$.   
The other half  will be full of 0's and 1's, the 1's being in the   
positions associated with that particular source, the 0's elsewhere.  
That is, the lens equation can be written explicitly as   
($\hat{a}$ denotes matrix and $\vec{a}$ denotes vector);  
\begin{equation}  
\left( \begin{array}{c} \vec{\theta_x} \\  
                        \vec{\theta_y} \end{array} \right) =   
\left( \begin{array}{ccc} \hat{\Upsilon_x} \ \ \hat{1} \ \ \hat{0} \\  
                          \hat{\Upsilon_y} \ \ \hat{0} \ \ \hat{1} \end{array} \right)  
\left( \begin{array}{c} \vec{M} \\  
                        \vec{\beta_o^x}\\  
                        \vec{\beta_o^y}\\ \end{array} \right)  
\label{eq_lens3}  
\end{equation}  
Where again, $\vec{\theta_x}$ and $\vec{\theta_y}$ are   
two $N_{\theta}$ dimensional vectors containing the x and y positions,   
respectively, of the pixels in the observed arcs.   
The two ($N_{\theta} \times N_c$ matrices $\hat{\Upsilon_x}$   
and $\hat{\Upsilon_y}$ contain the x and y lensing effect of the cell $j$ on the   
$\theta$ pixel $i$. The $N_{\theta} \times N_s$ dimensional matrices $\hat{1}$ 
and $\hat{0}$ are full of 0's (the $\hat{0}$ matrix) or contain 1's   
(the $\hat{1}$ matrix) in the $i$ positions ($i \in [1,N_{\theta}]$)   
where the $i$th $\theta$ pixel comes from the $j$   
source ($j \in [1,N_s]$) and 0 elsewhere.   
The vector $\vec{M}$ contains the $N_c$ gridded masses we want to estimate.   
 $\vec{\beta_o^x}$ contains the central x positions of the $N_s$ sources.   
Similarly, $\vec{\beta_o^y}$ will contain the central y positions.  
  
%%%%%%%%%%%%%%%%%%%%%%%%%%%%%%%%%%%%%%%%%%%%%%%%%%%%%%%%%%%%%%%%%%%%%%%  
  
%%%%%%%%%%%%%%%%%%%%%%%%%%%%%%%%%%%%%%%%%%%%%%%%%%%%%%%%%%%%%%%%%%%%%%%  

\bsp  
\label{lastpage}  
\end{document}